# Hilbert phase microscopy based on pseudo thermal illumination in Linnik configuration

MIKOŁAJ ROGALSKI,[1,4] MARIA CYWIŃSKA,[1,4] AZEEM AHMAD,[2] KRZYSZTOF PATORSKI,[1] VICENTE MICÓ,[3] BALPREET S. AHLUWALIA[2], MACIEJ TRUSIAK[1,*]

[1]*Warsaw University of Technology, Institute of Micromechanics and Photonics, 8 Sw. A. Boboli St., 02-525 Warsaw, Poland*
[2]*Departament of Physics and Technology, UiT The Arctic University of Norway, NO-9037 Tromsø, Norway*
[3]*Departamento de Óptica y de Optometría y Ciencias de la Visión, Universitat de Valencia, C/Doctor Moliner 50, Burjassot 46100, Spain*
[4]*Authors contributed equally to this work.*
[*]*maciej.trusiak@pw.edu.pl*

**Quantitative phase microscopy (QPM) is often based on recording an object-reference interference pattern and its further phase demodulation. We propose Pseudo Hilbert Phase Microscopy (PHPM) where we combine pseudo thermal light source illumination and Hilbert spiral transform phase demodulation to achieve hybrid hardware-software-driven noise robustness and increase in resolution of single-shot coherent QPM. Those advantageous features stem from physically altering the laser spatial coherence and numerically restoring spectrally overlapped object spatial frequencies. Capabilities of the PHPM are demonstrated analyzing calibrated phase targets and live HeLa cells in comparison with laser illumination and phase demodulation via temporal phase shifting and Fourier transform techniques. Performed studies verified unique ability of the PHPM to couple single-shot imaging, noise minimization, and preservation of phase details.**

Quantitative phase imaging (QPI) [1-3] enhances optical microscopy as it offers high contrast label-free observation of unimpaired biological cells and tissues via phase delay distribution measurements (related to refractive index and/or sample thickness variation). Sample induced refraction is coded in the phase part of complex optical field. Incoherent light based multi-frame phase retrieval methods [4] efficiently cope with phase noises, however nonlinear solvers tend to distort very high and very low object spatial frequencies. Quantitative phase measurements based on coherent laser light illumination and principles of interferometry and holography [1-3] employ straightforward deterministic phase demodulation from recorded spatial interference image called fringe pattern. They are affected by coherent noises and artifacts, however, and tend to limit high object spatial frequencies in single-shot operation mode (indispensable in dynamic object investigation). Phase can be accurately demodulated using multi-hologram approaches based on temporal phase shifting (TPS) [3] with full usage of the detector bandwidth. In TPS, limited imaging speed constitutes the trade-off. It is therefore of high importance to establish a QPI technique able to provide single-shot imaging capability with minimized phase noise and maximized phase space-bandwidth product [5] (PSBP, information capacity measure).

Low temporal coherence sources have been used to remove coherent artifacts from the interference fringe pattern [6-8], however short coherence length often results in limiting accessible optical path difference (OPD) range and damaging the contrast of interference fringes. Pseudo-thermal light source (PTLS), one of the most important examples of spatial coherence alteration in QPI [9]. Its high temporal and low spatial coherence enabled bypassing the OPD limitation and allowed to capture low noise interferograms over wide field-of-views [10-11]; similarly to the dynamic speckle illumination technique [12-13], however without the need to introduce a diffraction grating. TPS was used to perform accurate phase demodulation [10-11], thus penalizing measurement speed. Fourier transform (FT) phase retrieval [1-3] needs high carrier and limited object spatial frequency spectrum to be able to filter in the interferogram's spectrum the coherent cross-correlation term (1st order) residing outside the central spectral term (0th order). PSBP can be thus restricted, which prevents full appreciation of single-shot coherent noise minimization. Slightly off-axis QPI approaches [14] somewhat address this limitation via two-shot modification.

In this contribution we propose Pseudo Hilbert Phase Microscopy (PHPM) as first, to the best of our knowledge, interference-pattern-based QPI technique able to provide very low phase noise and very highly detailed phase content in a single camera exposure for fringes with overlapping spectrum (of low carrier frequency and coming from high numerical aperture). It employs PTLS to generate low noise interferogram and then embraces Hilbert spiral transform (HST) phase demodulation [15,16]. HST, working with background-free fringe pattern and fringe direction map, as fully 2D method advances well-established 1D HT [17]. PHPM enables to keep low magnification and preserve high resolution provided by the detector bandwidth. Typically, to achieve single-shot phase imaging, small filter size (limited phase resolution) Fourier transform phase retrieval is used after increasing the magnification (decreasing complex pupil's diameter) and adjusting fringe carrier frequency to avoid aliasing [3]. In other words, our method is designed to work when object frequency component (first spectral order) cannot be fully isolated in the hologram's Fourier space.

It is worth mentioning that recently proposed Kramers-Kronig (KK) phase retrieval scheme [5] loosens mentioned FT spectral restrictions and allows for cross-correlation term to overlap with autocorrelation term. The PSBP, although significantly increased, is at the same time still limited as KK requires separation of conjugate cross-correlation spectral terms, like in the slightly off-axis QPI scheme [14]. Although the HST is free of such caveats, it is heavily dependent on the quality of image processing algorithms used for fringe pattern incoherent background term removal and fringe direction map calculation. Empirical [16], variational [18] and iterative [19] algorithms recently emerged as very capable fringe pattern preprocessing methods (detrending and denoising).

The proposed PHPM can be divided into two parts: (1) hardware related and physics-driven wide-field coherent noise minimization via PTLS deployment and (2) software related high PSBP single-shot phase retrieval using HST. We describe both parts and discuss the novelty behind combined hardware-software approach. High quality phase imaging is corroborated using static phase targets – Siemens star, Fig. 2, and cross, Fig. 3 – with new quantitative way of investigating the phase transfer function as a standard deviation (STD) of the azimuthal star profile. Living HeLa cells constitute biological sample used for experimental verification of PHPM, Fig. 4. The PTLS is generated when a highly coherent laser beam hits a rotating diffuser. The output of the diffuser acts like an extended source with high monochromaticity. Such source brings various advantages in the QPM system such as high spatial phase sensitivity, high temporal resolution with large field of view (FOV) [10]. This type of light source also allows the extended range of OPD adjustment unlike broadband light sources and enables the use of non-identical objective lenses in the object and the reference arm of the QPM system [10,11]. Speckle noise is significantly limited, whereas shot noise in the interferogram is not an issue since we work in high photon budget regime. Figure 1 demonstrates optical setup implemented to record interferograms used throughout this study. The optical configuration is based on the Linnik interference microscopy system. The laser light beam coming from a coherent laser light source is coupled into a single mode fiber (SMF) using a condenser lens ($CL_1$). The output of SMF, which is a spatially filtered high coherent light beam, is coupled to the input port of the QPM system. The laser light beam is also directed towards the rotating diffuser (RD) by employing a flip mirror (FM) to generate PTLS. The output of RD is coupled into a multimode fiber (MMF) of core diameter of 1 mm using $CL_2$. The beam splitter (BS1) combined/directed the outputs of SMF and MMF towards the QPM system as illustrated in Fig. 1. The image of light source (output facet of the MMF) is relayed to the back focal plane of the microscope objective ($MO_1$) using lenses $L_1$ and $L_2$ to generate uniform illumination at the sample plane. Lenses $L_1$ and $L_2$ also focused the light beam coming from SMF at the back focal plane of $MO_1$ to generate collimated beam at the output of $MO_1$. $BS_2$ splits the beam into two: the object and reference beams are directed towards the sample and the reference mirror (M), respectively. The reflected light beams from the sample and M are recombined at the same beam splitter BS2 and superimposed at the camera using a tube lens to form interference pattern. In the reference arm, a fixed 10x/0.25 objective lens is used, whereas, in the object arm various objective lenses ($MO_2$) of different magnifications are used to achieve user defined resolution and magnification in QPM system. The reference mirror M is attached to a piezo electric transducer (PZT) stage to introduce nanometric temporal phase shift in the interferogram. The unit of piezo stage, $MO_1$ and M is attached to a long travel range motorized stage, which controlled the wavefront curvature of the reference beam and helped to match with the object beam wavefront curvature for different sample arm objective lenses.

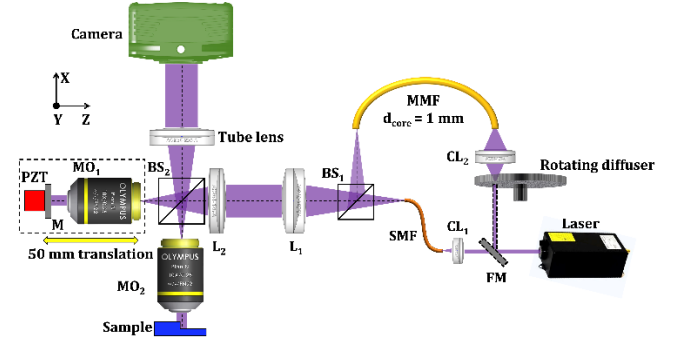

Fig. 1. Schematic diagram of quantitative phase microscope (QPM) in reflective configuration where two layouts (proposed PHPM and conventional TPS) share the same optical path and the way to select them is with the flip mirror (FM).

Omitting the noise presence for simplicity, the HST input signal (background-free fringe pattern) can be described as:
$$s(x,y) = b(x,y)cos(\varphi(x,y)), \qquad (1)$$
and then output signal follows as:
$$s_H(x,y) = -b(x,y)sin(\varphi(x,y)). \qquad (2)$$
Finally, the phase function can be calculated as:
$$\varphi(x,y) = \tan^{-1}\left(\frac{s_H(x,y)}{s(x,y)}\right). \qquad (3)$$

Using the HST nomenclature [21-23] the quadrature function can be described as:
$$s_H(x,y) = -iexp[-\beta(x,y)]F^{-1}\{S(u,v)F\{s(x,y)\}\}, \qquad (4)$$
where b(x,y) denotes amplitude term, $F$ is a Fourier transform, $F^{-1}$ is an inverse Fourier transform, $S(u,v)$ is spiral phase function in spatial frequencies $(u,v)$, and $\beta(x,y)$ is local fringe direction map.

Minimizing coherent noise via PTLS naturally increases the signal-to-noise ratio (SNR) of crucial importance in phase demodulation as intensity noise transfers into phase domain seamlessly. Especially in the HST, as it is not employing any spectral filtering (in contrary to heavily filtered FT) nor multi-frame averaging (in contrary to multi-frame TPS). The HST itself is not performing any denoising nor background removal as it employs spiral phase function in Fourier space to generate quadrature component to the investigated fringe pattern. Fringe direction map (modulo 2pi) needs to be provided to correct for local phase sign errors [18,19]. Phase unwrapping completes the data analysis path. It is of high importance to remove noise while preserving the cosine profile of the fringe pattern. Numerical methods showed promising results [16,18,19], however low SNR interferogram should be analyzed using at least two π-phase-shifted fringe patterns [20]. Upon their subtraction a higher SNR difference-interferogram is obtained, which is easier to analyze via HST demodulation. As single-shot high PSBP phase demodulation of low SNR data remains challenging we propose a novel, to the best of our knowledge, PHPM approach.

The choice of fringe prefiltration algorithm is important, and to show its effect we will preprocess fringe patterns with two capable techniques – unsupervised variational image decomposition uVID [18], Fig. 2, and 2D fringe pattern fast iterative filtering (fpFIF2) [19], Figs. 3 and 4. The uVID preserves fringe details, however computations are quite burdened by iterative Chambolle projection algorithm. The fpFIF2 is advantageous due to its fast processing and fringe profile preservation, however for very complicated fringes precise uVID calculations allow for error-free phase unwrapping.

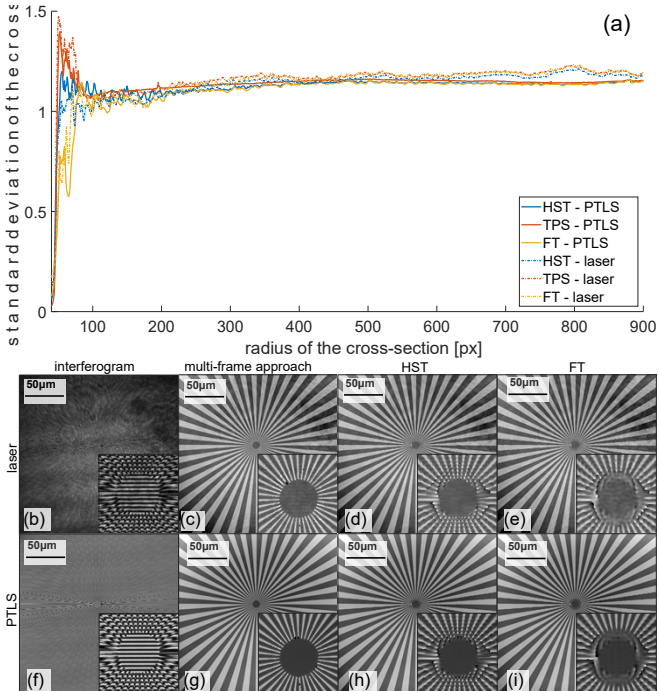

Fig. 2. Plots with radial STD analysis (a); interferograms (b,f) and phase maps obtained by TPS (c,g), HST (d,h) and FT (e,i) for laser (b-e) and PTLS (f-i) illumination. Visualization 1 shows dynamic cross-sections.

Figure 2 presents the case of Siemens star phase imaging with 10x/0.25 objective lens in the sample arm. Using multi-frame phase demodulation [21] a ground truth phase map was obtained for both laser (Fig. 2(c)) and PTLS (Fig. 2(g)) illumination. We then analyzed single interferograms from laser (Fig. 2(b)) and PTLS (Fig. 2(f)) series employing FT (Figs. 2(e) and 2(i)) and HST (Figs. 2(d) and 2(h)) methods. We can clearly observe the advantage of higher PSBP achievable via HST demodulation and the noise limitation accomplished applying PTLS. To quantify how favorably HST compares to the FT we decided to calculate standard deviation for circular cross-sections with increasing radii. Fig. 2(a) presents 3 plots for TPS/HST/FT demodulation under laser illumination (dashed lines) and 3 plots for TPS/HST/FT for PTLS (solid lines). The noise influence in the case of laser illumination is clearly visible since in the case of higher cross-section radii (with star features of lower frequency) the STD curves are not as stable as in the case of PTLS illumination. In the radii range of high frequency star features the HST curves fit better to the reference TPS curves than FT ones. It means that HST was able to catch the high frequency details, which were not available for FT approach. Visualization 1 helps to grasp the working principle of such quantitative evaluation and presents how under such challenging conditions HST demodulation provided similar results to TPS and surpassed FT method (for both PTLS and laser). Generated results verify PHPM (HST+PTLS) as favorable high-PSBP single-shot robust QPI method.

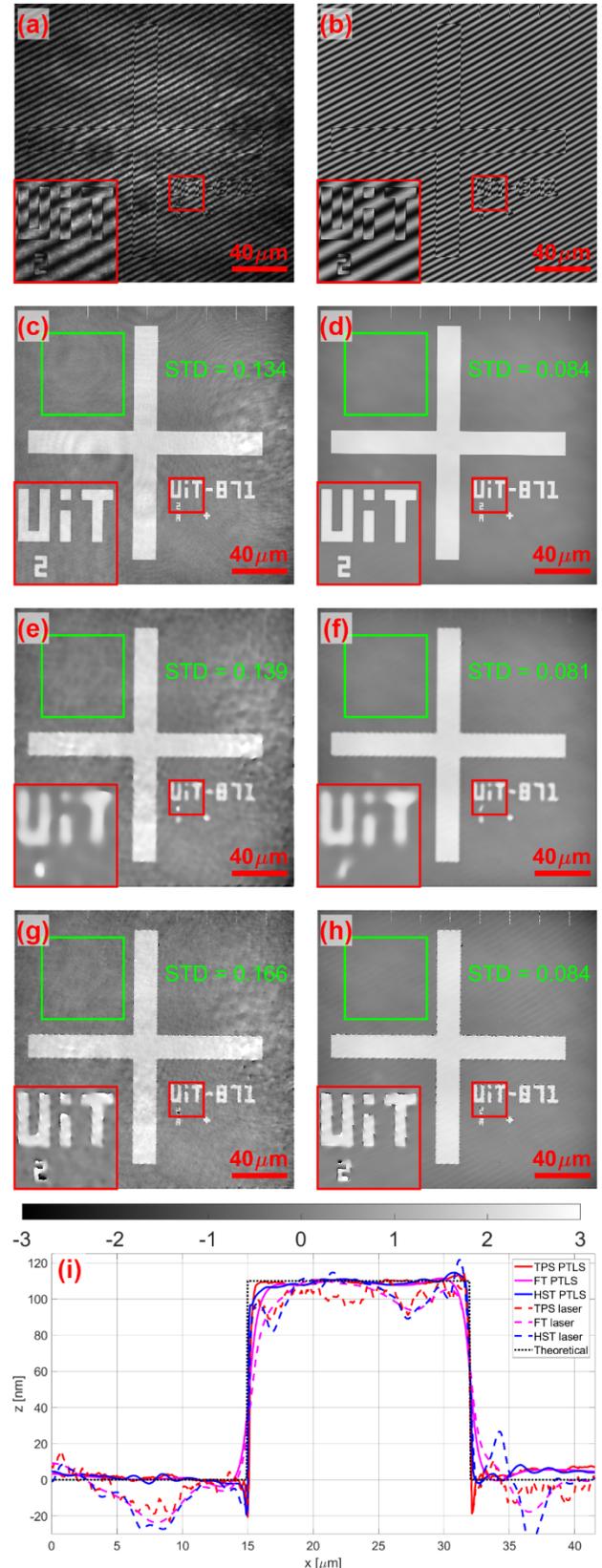

Fig. 3. Phase test target analysis: laser (a) and PTLS (b) interferograms; TPS (c, d), FT (e, f) and HST (g, h) phase maps; (c, e, g) were obtained for laser and (d, f, h) for PTLS. Cross-sections through 110 nm step (i).

Figure 3 presents the case of customized cross test (110 nm height) phase-target imaging. Figure 3(a) depicts a fringe pattern collected with the laser light source and Figs. 3(c), 3(e) and 3(g) present reconstructed phases for regular laser illumination with TPS, FT and HST, respectively. Fringe pattern recorded under PTLS illumination is shown in Fig. 3(b) and phases reconstructed with TPS, FT and HST are presented in Figs. 3(d), 3(f) and 3(h), respectively. For each phase, we calculated the STD (in radians) within an object-free area marked with green rectangle (values are shown in Fig. 3), which may be treated as a measure of noise presence in the reconstruction [22,23]. As can be observed, both, fringe pattern and reconstructed phases are less noisy for PTLS, which is confirmed by calculated STD values. Interestingly, PTLS removed mainly medium and high frequency speckles, while low-frequency speckle component is a main factor determining STD values. It can be also seen that single-frame FT significantly lowers the reconstruction resolution compared to multi-frame TPS (see enlarged regions in Fig. 3). The PHPM reconstruction, although slightly spoiled with some phase-jump-related artifacts, provides TPS-like phase accuracy and resolution (see cross-section in Fig. 3(i)) - both significantly higher than the FT outcomes.

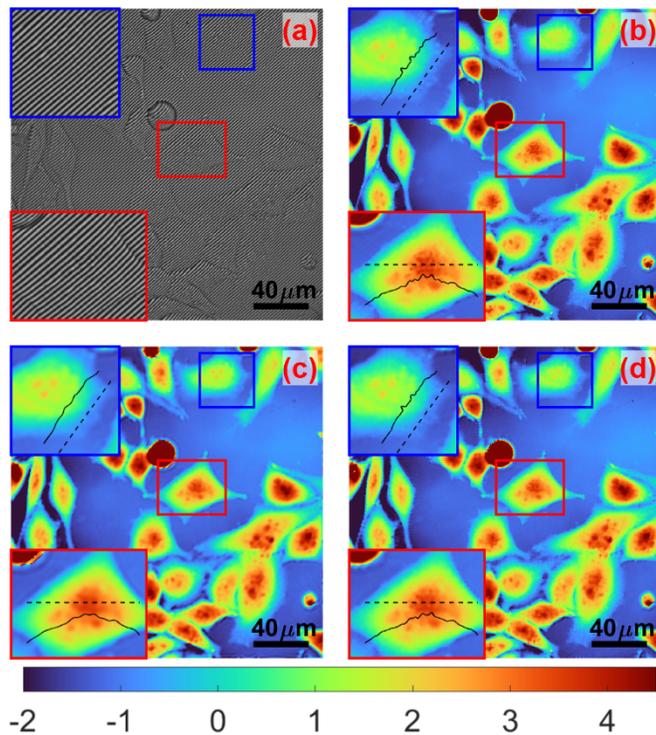

Fig. 4. Fringe pattern acquired with the use of PTLS light source (a) and phases reconstructed with TPS (b), FT (c), and HST (d). Visualization 2 shows the dynamic phase imaging of live HeLa cells.

To further investigate the single-shot PHPM capabilities and demonstrate its applicability to solving real-life QPI problems we have tested it in regime of living HeLa cells imaging, Fig. 4 and Visualization 2, using water immersion objective lens (60x/1.2NA) in the sample arm of QPM. Collected fringe pattern with PTLS illumination is shown in Fig. 4(a) and phases reconstructed with TPS, FT and HST are presented in Figs. 4(b), 4(c) and 4(d) respectively. Comparing to FT, the single-shot PHPM reconstruction achieves significantly higher resolution, which is comparable to multi-frame TPS method (cross-sections in Figs. 4(b), 4(c) and 4(d)). In conclusion, we proposed the Pseudo Hilbert Phase Microscopy, a first, to the best of our knowledge, single-shot QPI technique able to simultaneously ensure very low phase noise and high phase resolution over a wide field of view without full separation of spectral sidelobes. This way PHPM offers phase imaging enhanced both in terms of signal-to-noise-ratio and resolution, in comparison with typically available QPI solutions. Attractive capabilities of PHPM were verified using static phase test targets examination and living HeLa cells imaging.

**Funding.** National Science Center Poland (2020/37/B/ST7/03629). INTPART (project: 309802), Research Council of Norway, project NANO 2021-288565 and BIOTEK 2021-285571. Grant PID2020-120056GB-C21 funded by MCIN/AEI/10.13039/501100011033.

**Disclosures.** The authors declare no conflicts of interest.

**Data availability.** Data underlying the results presented in this paper are not publicly available at this time but may be obtained from the authors upon reasonable request.